\documentclass[reprint,superscriptaddress,amsmath,amssymb,aps,prb]{revtex4-1}
\usepackage{graphicx,color}
\usepackage{amssymb}   % for math
\usepackage{amsmath}
\usepackage{amsfonts}
\usepackage{mathdots}
\usepackage{hyperref}
\usepackage{epsfig}
\usepackage{hyperref}
\usepackage{epstopdf}
\usepackage{bm}
\usepackage{braket}
\begin{document}
\title{Strongly enlarged topological regime and enhanced superconducting gap in nanowires coupled to Ising superconductors}
\author{ Ying-Ming Xie}
\author{ Benjamin T. Zhou}
\author{T. K. Ng}
\author{K. T. Law} \thanks{phlaw@ust.hk}
\affiliation{Department of Physics, Hong Kong University of Science and Technology, Clear Water Bay, Hong Kong, China. }
\begin{abstract}
An external magnetic field is needed to drive a nanowire in proximity to an $s$-wave superconductor into a topological regime which supports Majorana end states. However, magnetic field generally suppresses the proximity superconducting gap induced on the nanowire. In recent experiments using InSb nanowires coupled to Al, the induced proximity gap vanishes at magnetic fields $B\sim 1$T. This results in a small superconducting gap on the wire and a narrow topological regime which is proportional to the strength of the magnetic field. In this work, we show that by placing nanowires in proximity to recently discovered Ising superconductors such as the atomically thin transition-metal dichalcogenide(TMD) NbSe$_2$, the topological superconducting gap on the wire can maintain at a large magnetic field as strong as $B \sim 10$T. This robust topological superconducting gap is induced by the unique equal-spin triplet Cooper pairs of the parent Ising superconductor. The strong magnetic field allows a topological regime ten times larger than those in InSb wires coupled to Al. Our work establishes a realistic platform for building robust Majorana-based qubits. 
\end{abstract}
\pacs{}
\maketitle

\section{Introduction}
Majorana zero modes (MZMs) are non-Abelian particles\cite{Ivanov2001} which can serve as potential building blocks of topological quantum computers\cite{Nayak}. It was first shown that MZMs can form at the vortex cores of two-dimensional $p$-wave superconductors\cite{Read}. Kitaev later pointed out that this zero energy mode appear at the ends of 1D spinless $p$-wave superconducting wire\cite{kitaev2001unpaired}. Despite the lack of $p$-wave superconductors in nature, effective $p$-wave pairing can be engineered in semiconductors in proximity to conventional $s$-wave superconductors\cite{Fu2008, Oreg2010,Jaysau2010,Choy2011,DanielLoss2013,Perge2014,Chuizhen2018}. In particular, promising signatures of MZMs have been been demonstrated experimentally in semiconducting InSb nanowires coupled to superconducting aluminum (Al) upon application of external magnetic fields\cite{Mourik2012,Marcus2016,Deng2016,Jia2016}. Recent experimental observation of the predicted quantized zero bias peak\cite{Zhang2018} has provided strong evidence for the existence of MZMs and motivated active on-going efforts toward their applications in topological quantum computations\cite{Alicea2016,Lutchyn2018}. 

To realize a practical topological quantum computer, a robust topological superconducting gap is essential for the integrity of Majorana-based qubits. However, achieving a hard-gap nanowire has so far been experimentally challenging\cite{zhang2017ballistic,Deng2016,Zhang2018}. One important reason is the dual character of magnetic fields in creating topological superconductivity: on one hand, an external magnetic field is needed to create a single species of fermions such that the system becomes an effective spinless Kitaev chain when pairing is introduced; on the other hand, the induced topological gap by usual singlet Cooper pairs gets easily suppressed upon increasing applied magnetic fields\cite{zhang2017ballistic,Deng2016,Zhang2018}. This puts a stringent constraint on the tunable topological regime of the nanowires, thus making the Majorana-based qubits susceptible to applied magnetic fields. A feasible way to engineer a robust topological gap against external fields is clearly desirable.

Recently, Ising superconductors with remarkably high in-plane upper critical fields $B_{c2} \sim 30-50$T have been discovered in several transition-metal dichalcogenides(TMDs) such as gated MoS$_2$ thin films\cite{Lu2015,Saito2015} and atomically thin NbSe$_2$\cite{Xi2015}. The strong enhancement of in-plane $B_{c2}$ was explained to arise from the special Ising spin-orbit coupling(SOC) of TMD materials, which pins electron spins at opposite momentum to opposite out-of-plane directions. Interestingly, with $s$-wave pairing potential only, the Ising SOC generates equal-spin triplet Cooper pairs in Ising superconductors, with their spins pointing to in-plane directions\cite{Law2016}. Under applied in-plane magnetic fields, these equal-spin Cooper pairs align their spin magnetic moments along the direction of applied in-plane fields, which lowers the free energy of Ising superconductors under in-plane fields and gives rise to strongly enhanced in-plane $B_{c2}$. 

Importantly, these equal-spin Cooper pairs can tunnel into nanowires under in-plane magnetic fields\cite{Law2016} or magnetic atomic chains\cite{Tewari2016,Aji2016} via proximity effects, which results in a Kitaev chain supporting MZMs. Given the compatibility between these equal-spin Cooper pairs and in-plane spin magnetization, a natural question follows whether the topological gap in nanowires created by these Cooper pairs can stay robust against strong in-plane Zeeman fields.

In this work, we show that the answer is positive: the proximity-induced topological gap by the equal-spin Cooper pairs in Ising superconductors is almost insensitive to in-plane Zeeman fields. This is in sharp contrast to the effective $p$-wave pairing created by $s$-wave superconductors, which can be easily suppressed upon application of external magnetic fields. 

In particular, by considering a realistic heterostructure formed by InSb nanowires/superconducting monolayer NbSe$_2$, we demonstrate that the topological superconducting gap induced by NbSe$_2$ can persist under applied fields as strong as $B \sim 10$T, which far exceeds the critical field strengths($B_{c}\sim 1$T) in previous experiments using superconducting aluminium\cite{Mourik2012}. Importantly, the large external magnetic field creates a large Zeeman gap between the spin polarized subbands of the InSb wire such that the topological regime, which is proportional to the Zeeman gap, is strongly enlarged. The large superconducting gap and the large topological regime provide a promising platform for realizing Majorana fermions in nanowire/Ising superconductor heterostructures.

In the following sections, we first discuss the important role of equal-spin Cooper pairs in enhancing the in-plane $B_{c2}$ in Ising superconductors. We explain how these equal-spin Cooper pairs can couple to in-plane external magnetic fields and make the superconducting state energetically favorable at in-plane fields far exceeding conventional Pauli limits. Second, by analytically projecting the contribution of an Ising superconductor onto proximity-coupled nanowires, we show that the equal-spin pairing induced by Ising superconductors is insensitive to in-plane Zeeman fields. Finally, using a realistic tight-binding model, we study the magnetic-field dependence of the topological superconducting gap in quasi-1D InSb nanowires placed next to a superconducting monolayer NbSe$_2$. We thus establish a realistic platform based on Ising superconductors for building robust Majorana-based qubits.  

%\section{SMN/ISC heterostructure}
\section{Enhancement of in-plane $B_{c2}$ due to equal-spin Cooper pairs}

In this section, we explain the role of equal-spin Cooper pairs in enhancing in-plane $B_{c2}$ in Ising superconductors. As we are about to show, these in-plane equal-spin Cooper pairs reconcile the competition between superconductivity and in-plane spin magnetization, which creates a robust superconducting order in Ising superconductors against in-plane fields.

In the following discussions, we ignore the orbital pair-breaking effects of in-plane fields. This simplification is valid given that the recently found Ising superconductors such as 2H-NbSe$_2$\cite{Xi2015} or 2H-MoS$_2$\cite{Lu2015,Saito2015} have atomic-scale thickness far smaller than the superconducting coherence length.

%In this section, we first present a general discussion on the pairing correlation and the pairing potential of a non-centrosymmetric superconductor in the presence of magnetic fields. Then, we study how an in-plane magnetic field will affect the pairing gap of superconductors with/without the Ising SOC. 

In general, the normal state of an Ising superconductor can be described by the following two-band model Hamiltonian:
\begin{equation}	
H_0=\sum_{\bm{k},s_1,s_2}c_{\bm{k}s_1}^{\dagger}[ \xi(\bm{\bm{k}})\delta_{s_1,s_2}+(\bm{g}(\bm{k})\cdot\bm{\sigma})_{s_1,s_2} ] c_{\bm{k}s_2}
\label{eq:Hn}
\end{equation}

where $\xi(\bm{k})=\frac{\hbar^2\bm{k}^2}{2m}-\mu$ is the kinetic energy term, $\bm{\sigma}$ operates on the spin space, the $\bm{g}(\bm{k})$ term represents generic non-centrosymmetric spin-orbit coupling(SOC) terms, with $\bm{g}(\bm{k})=-\bm{g}(-\bm{k})$ imposed by time reversal symmetry. For intrinsic Ising superconductors such as monolayer NbSe$_2$, the $D_{3h}$ point group symmetry of 2H-TMDs dictates that the SOC terms can only pin electron spins to the out-of-plane directions, \textit{i.e.}, $\bm{g}(\bm{k}) = (0,0,\beta(\bm{k}))$, thus referred to as the Ising SOC. The effect of Zeeman fields is included by rewriting $\bm{g}(\bm{k})$ as $\bm{g}(\bm{k},\bm{V})=\bm{g}(\bm{k})+\bm{V}$, where $\bm{V}$ is given by the product between the Bohr magneton $u_B$ and the applied magnetic field $\bm{B}$. Due to the presence of Ising SOCs, the normal state spectrum exhibits a band splitting given by $\xi_{\pm}(\bm{k})=\xi(\bm{k})\pm|\bm{g}(\bm{k})|$. 

%The general interacting Hamiltonian for superconducting pairing has the form:
%\begin{equation}
%H_{int}=\frac{1}{2\Omega}\sum_{\bm{k},\bm{k'}}\sum_{s_1,s_2,s_1',s_2'}V_{s_1s_2s_2's_1'}(\bm{k},\bm{k'})c^{\dagger}_{\bm{k}s_1}c^{\dagger}_{-\bm{k}s_2}c_{-\bm{k}'s_2'}c_{\bm{k}'s_1'}
%\end{equation}
For the superconducting state, we assume the dominant pairing channel is the on-site attractive interaction, and the mean-field pairing potential is simply given by:
\begin{equation}
H_{pair}=\sum_{\bm{k}}\sum_{s_1,s_2} \Delta (i \sigma_{y})_{s_1, s_2} c^{\dagger}_{\bm{k}s_1}c^{\dagger}_{-\bm{k}s_2} + h.c.
\label{eq:Hpair}
\end{equation}
where $\Delta$ is the $s$-wave order parameter. As pointed out by previous works\cite{PAFrigeri}, the spin structure of Cooper pairs in a superconductor with SOCs can be conveniently described by the pairing correlation defined as:
\begin{equation}
F_{\alpha\beta}(\bm{k},\tau_1;\bm{k'},\tau_2)=\left\langle T_{\tau}c_{\bm{k},\alpha}(\tau_1)c_{\bm{-k'},\beta}(\tau_2)\right\rangle.
\label{eq:PairCorr}
\end{equation}
In the Matsubara frequency space, the pairing correlation $F$ can be written as a compact matrix form\cite{PAFrigeri}:
\begin{equation}
F(\bm{k},i\omega_n) =(F_s(\bm{k},i\omega_n)+\bm{F}_t(\bm{k},i\omega_n)\cdot\bm{\sigma})\Delta i\sigma_y,
\end{equation}
where $F_s$/$\bm{F}_t$ parametrize the singlet/triplet correlation functions respectively. By solving the Gor'kov equations (see Appendix A for details), $F_s$ and $\bm{F}_t$ can be obtained as:
\begin{align}
&F_s(\bm{k},i\omega_n)=\frac{1}{2}(\frac{1}{\varphi_{+}(\bm{k},\Delta,\omega_n)}+\frac{1}{\varphi_{-}(\bm{k},\Delta,\omega_n)})\label{spinsingletcorr},\\
&\bm{F}_{t}(\bm{k},i\omega_n)=\frac{1}{2}(\frac{1}{\varphi_{+}(\bm{k},\Delta,\omega_n)}-\frac{1}{\varphi_{-}(\bm{k},\Delta,\omega_n)})\hat{\bm{g}}(\bm{k}),
\label{eq:PairCorrMatrix}
\end{align}
with the specific forms of $\varphi_{\pm}$ given in Appendix A. Notably, due to the presence of SOCs, there is a nonzero triplet pairing correlation even though the mean-field potential is $s$-wave\cite{Rashba2001,Mackenzie2003,Law2016}. In particular, the triplet pairing correlation is parametrized by a vector-valued function $\bm{F}_{t}$. For any fixed $\bm{k}$, $\bm{F}_{t}$ is parallel to the SOC vector $\bm{g}$ according to Eq.\ref{eq:PairCorrMatrix}. 

\begin{figure}
	\centering
	\includegraphics[width=0.49\textwidth]{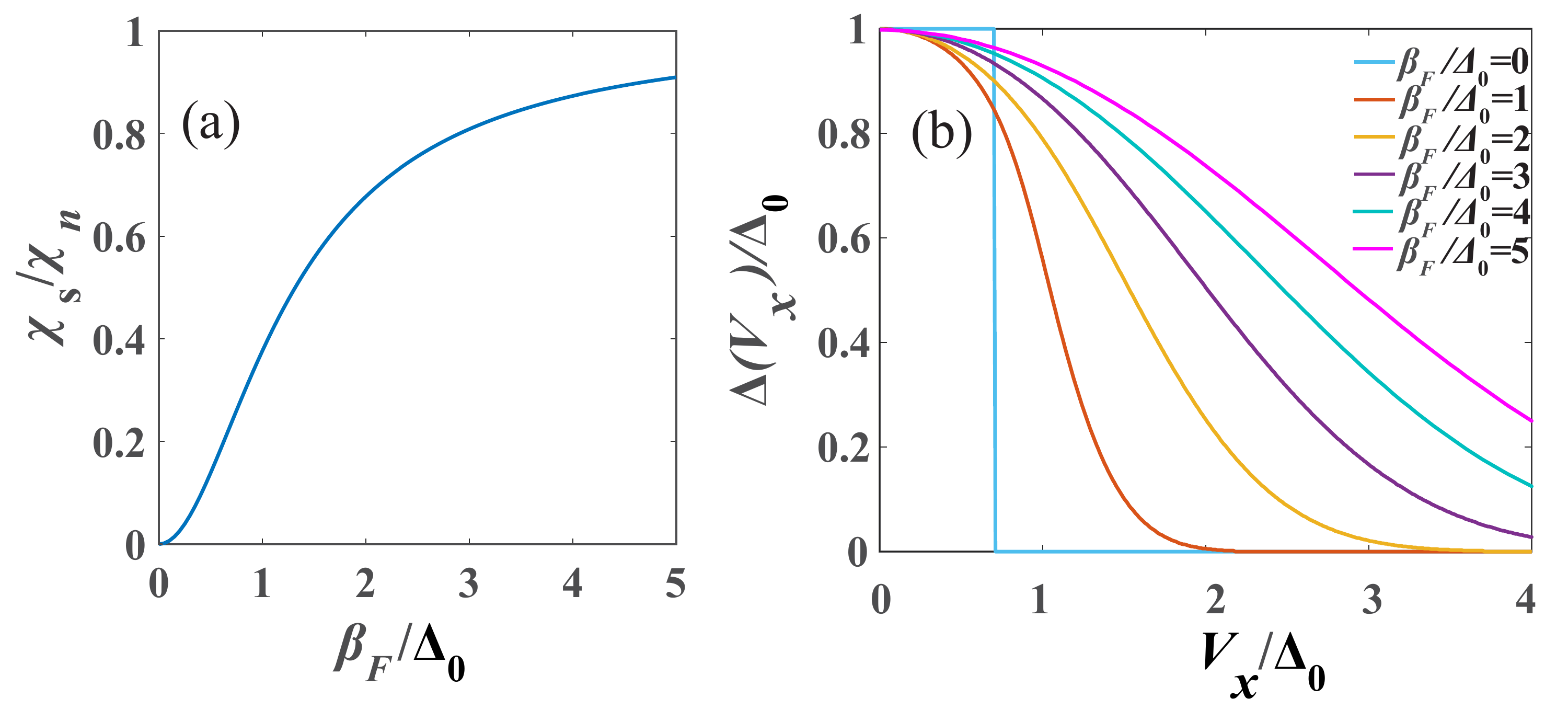}
	\caption{(a) The superconducting spin susceptibility $\chi_s$ as a function of Ising SOC strength $\beta_F$ at the Fermi energy. Without Ising SOC($\beta_F = 0$), $\chi_s=0$. As Ising SOC is turned on, spin triplet Cooper pairs start to form in Ising superconductors and gives rise to a nonzero $\chi_s$. (b) Superconducting gap $\Delta$ obtained self-consistently(Eq.\ref{selfconsistent}) at zero temperature as a function of Zeeman energy $V_x = \mu_B B_x$ under different Ising SOC strengths.  When $\beta_F=0$, superconductivity is destroyed at the usual Pauli limit: $B_{p}=\Delta_0/(\sqrt{2}\mu_B)$. When the strength of Ising SOC is finite, the upper critical field is enhanced, which exceeds $B_{p}$ by several times. }
	\label{fig:gapv}
\end{figure}

In the specific case of Ising superconductors, the unit SOC vector $\hat{\bm{g}} = \hat{z}$, which implies that the triplet correlation $\bm{F}_{t}$ has a nonzero $z$-component only. Under the basis of out-of-plane spin states (the out-of-plane $z$-axis defined as the spin quantization axis), Eq.\ref{eq:PairCorrMatrix} suggests that the triplet Cooper pairs are formed by electrons of opposite out-of-plane spins, \textit{i.e.}, the spinor wavefunction is given by: $\ket{\uparrow \downarrow} + \ket{\downarrow \uparrow}$. Interestingly, by a  straightforward change of basis from out-of-plane to in-plane spins, these triplet Cooper pairs have their spinor part given by equal-spin configurations. Without loss of generality, by defining the $x$-axis as the spin quantization axis, the triplet state becomes: $\ket{\uparrow \downarrow} + \ket{\downarrow \uparrow} \equiv \ket{\rightarrow \rightarrow} - \ket{\leftarrow \leftarrow}$. Thus, the spinor wavefunction of these triplet Cooper pairs is an equal superposition of spinor states with opposite in-plane spin magnetic moments. Under applied in-plane fields $B_x$, these in-plane equal-spin Cooper pairs can align their spin magnetic moments along the magnetic field direction. This endows the Ising superconductor with a finite spin susceptibility as shown in Fig.\ref{fig:gapv}(a). As a result, an Ising superconductor gains magnetic energy under in-plane magnetic fields, with its superconducting free energy kept lowering as the field strength increases. This leads to an enhanced in-plane $B_{c2}$.

To demonstrate the enhancement of $B_{c2}$ due to Ising SOC, we solve the superconducting gap self-consistently as a function of in-plane magnetic field. Without loss of generality, we assume $\bm{V} = V_x \hat{x}$, and the self-consistent gap equation is given by
 \begin{equation}\label{eq:GapEqn}
 \Delta i\sigma_y=T\sum_{\bm{k},n}V_0F_s(\bm{k},V_x,i\omega_n)\Delta i\sigma_y,
 \end{equation}  
where $V_0$ is the interaction strength, $F_s(\bm{k},V_x,i\omega_n)$ is the pairing correlation including the magnetic field. Details of self-consistent gap equations and spin susceptibilities are presented in Appendix \ref{Ap_Anomalous Green}-\ref{ApA}. 

In the zero-temperature limit, the superconducting gaps as a function of $V_x = \mu_B B_x$ with different Ising SOC strengths are shown in Fig.\ref{fig:gapv}(b). Without Ising SOCs, superconductivity is destroyed by the paramagnetic effect at the Pauli limit $B_{p}\approx \Delta_0/\sqrt{2}\mu_B$ (light blue curve in Fig.\ref{fig:gapv}(b)). The superconductor undergoes a first-order phase transition with its order parameter vanishing abruptly to zero. In contrast, in the presence of Ising SOCs, the upper critical magnetic fields are strongly enhanced to several times of the usual Pauli limit. Moreover, the superconducting gap decreases gradually to zero as $V_x$ increases, signifying a continuous phase transition at $B_{c2}$. This continuous superconductor-normal phase transition is recently observed experimentally in atomically thin NbSe$_2$ \cite{Fai2018continue}, which confirms the robust superconducting order of Ising superconductors against in-plane magnetic fields.

% When the Ising SOC energy is very strong which can be dozens of times of pairing gap for monolayer superconducting TMD 2H-MoS$_2$ or 2H-NbSe$_2$ in experiments\cite{Lu2015,Xi}, the spin is locked along the out of plane direction. In this case, the critical in-plane magnetic field can be even higher than 30T\cite{Xi}. 

%So the motivation of this work is to think about what will happen if a semiconductor nanaowire is in proximity to this Ising superconducting background. Will the critical field that the topological superconducting gap can survive be increased or how the topological superconducting gap behaves with the increasing of magnetic field? After the explicit calculation in the next few parts, we really find the topological superconducting gap can survive to much stronger external magnetic field region due to  the equal spin cooper paring forming by the Ising superconductor background.

\section{Robust topological superconductivity in nanowires coupled to Ising superconductors}
In this section, we study in details the superconducting proximity effects in semiconducting nanowires(NWs), such as InSb NWs, placed on top of an Ising superconductor. In particular, we demonstrate how a robust topological gap can be created in InSb NWs using the equal-spin Cooper pairs in Ising superconductors. 

\begin{figure}
	\centering
	\includegraphics[width=3.5in]{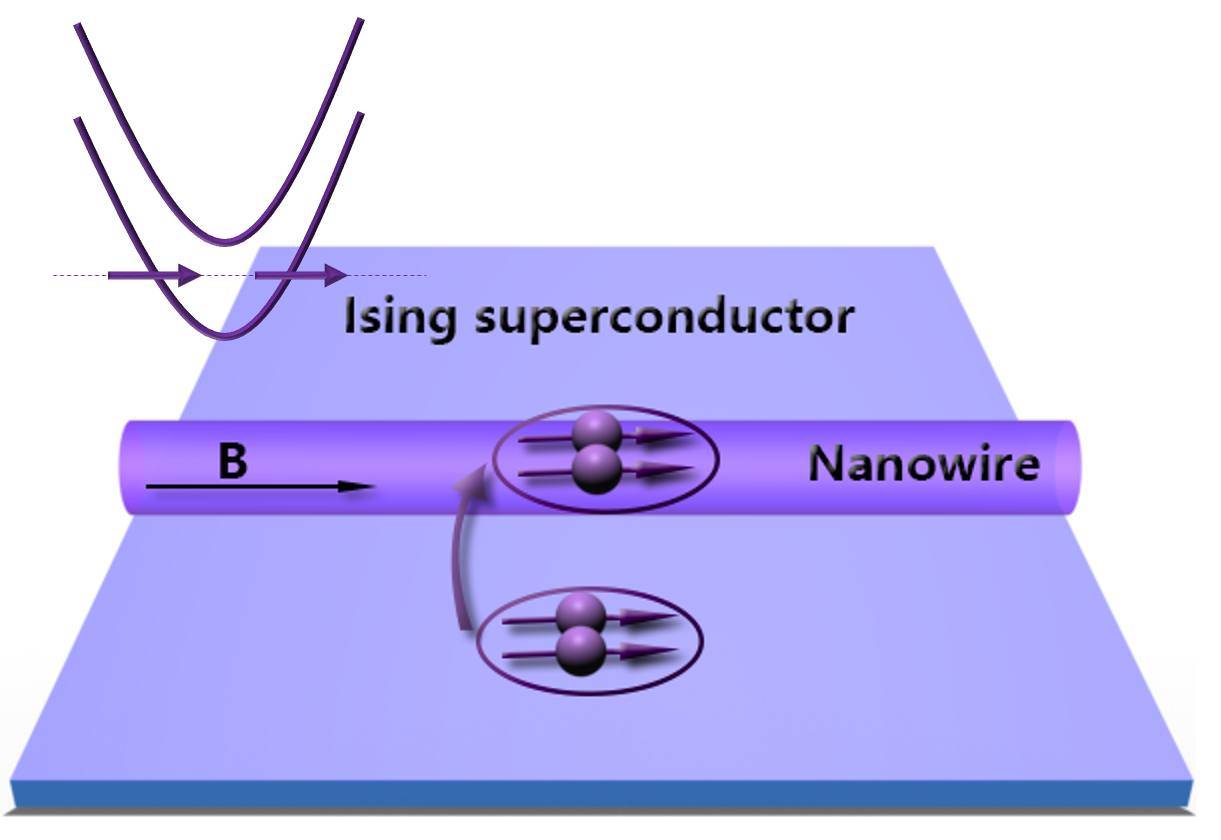}
	\caption{Schematic of a nanowire in proximity to an Ising superconductor. $B$ is the in-plane magnetic field required to induce sizable spin-splitting such that a single band is filled at the Fermi energy. The nanowire acquires a proximity gap from the equal-spin Cooper pairs in Ising superconductors and becomes topologically equivalent to a Kitaev chain. }
	\label{fig:nanotmd}
\end{figure}

In heterostructures formed by InSb NWs and conventional superconductors such as aluminum (Al), the proximity-induced gap originates from the parent $s$-wave superconducting gap, which can be easily suppressed under strong magnetic fields\cite{Oreg2010,Jaysau2010}. However, the situation can be very different using an Ising superconductor. First, as we demonstrated in the previous section, the parent superconducting gap of Ising superconductors is much less sensitive to external in-plane fields. Second, the special equal-spin Cooper pairs from Ising superconductors, which are compatible with in-plane fields, can tunnel into InSb nanowires via proximity effect as depicted in Fig.\ref{fig:nanotmd}. Expectedly, they can create a robust proximity-gap which stays robust at much higher in-plane fields.

\subsection{Strictly one-dimensional nanowire on a generic Ising superductor}\label{PartB}
To study the proximity-induced superconducting pairing, we first consider a strictly one-dimensional InSb nanaowire placed on a generic Ising superconductor described by the simple two-band model presented in the previous section.
The Hamiltonian of Ising superconductors in Nambu basis $(\psi^{(s)}_{\bm{k}\uparrow},\psi^{(s)}_{\bm{k}\downarrow},\psi^{(s)\dagger}_{-\bm{k}\uparrow}, \psi^{\dagger(s)}_{-\bm{k}\downarrow})^{T}$ is 
\begin{equation}
H_s=\sum_{\bm{k}}\psi^{\dagger(s)}(\bm{k})\mathcal{H}_{BDG}(\bm{k})\psi^{(s)}(\bm{k}),
\end{equation}
where 
\begin{equation}
\mathcal{H}_{BdG}(\bm{k})=\begin{pmatrix}
H_0(\bm{k})&\Delta i\sigma_y\\
(\Delta i\sigma_y)^{\dagger}&-H_0^T(-\bm{k})
\end{pmatrix}.
\end{equation}
In the basis $[ c^{(w)}_{\uparrow}(x),c^{(w)}_{\downarrow}(x),c^{(w)\dagger}_{\uparrow}(x),c^{(w)\dagger}_{\downarrow}(x))] ^{T} $, the nanowire under in-plane magnetic field can be described by
\begin{eqnarray}
H_{w}&=&\int dxc^{(w)\dagger}(x)[(\frac{\hbar^2\hat{k}_x^2}{2m}-\mu_w)\tau_z+\alpha_R\hat{k}_x\sigma_y\tau_z\\\nonumber
&+&V_x\sigma_x\tau_z]c^{(w)}(x).
\end{eqnarray}
The tunneling Hamiltonian is
\begin{equation}
H_{s-w}=\int dx \psi^{(s)\dagger}(x)\mathcal{T}c^{(w)}(x)+h.c.,
\end{equation}
where $\mathcal{T}=\Gamma_c \tau_z$, $\Gamma_c$ is the coupling strength, $\tau_z$ operates on particle-hole space, $\mu_w$ is the chemical potential of the nanowire, $\alpha_R$ is the Rashba SOC strength.

The contribution from the Ising superconductor is included as a self-energy term:
\begin{align}
\Sigma(x,x';i\omega)&=\int \frac{dk_xd\bm{k}_{\perp}}{(2\pi)^2}\mathcal{T}^{\dagger}\mathcal{G}(\bm{k},i\omega)\mathcal{T}e^{-ik_x(x-x')}\\
&=\int \frac{dk_x}{2\pi}\Sigma(k_x;i\omega)e^{-ik_x(x-x')},
\end{align}
where $\Sigma(k_x;i\omega)=\mathcal{T}^{\dagger}\mathcal{G}^{Sur}(\bm{k},i\omega)\mathcal{T}$, $\mathcal{G}^{Sur}(\bm{k},i\omega)=\int \frac{d\bm{k}_{\perp}}{2\pi}\mathcal{G}(\bm{k},i\omega) $. As shown in Appendix \ref{ApB}, including $\Sigma$, the low energy effective Hamiltonian of the nanowires is written as
\begin{equation}
\hat{H}_{eff}(k_x)=
\begin{pmatrix}
\tilde{h}(k_{x})&\tilde{\Delta}(k_x)\\
\tilde{\Delta}(k_x)^{\dagger}&-\tilde{h}^{T}(-k_{x})
\end{pmatrix},
\end{equation}
where $\tilde{\Delta}(k_x)=
(\psi(k_{x})+\bm{d}(k_{x})\cdot\bm{\sigma})i\sigma_y$ denotes the proximity-induced pairing matrix. Details of $\tilde{h}(k_x),\psi(k_x),\bm{d}(k_x)$ are presented in Appendix \ref{ApB}. 

Now we study the components in $\tilde{\Delta}(k_{x})$ which can give rise to topological superconductivity. Considering the case in which the magnetic field is parallel to the nanowire, with its strength large enough such that only a single band in the wire is occupied at the Fermi energy. To see the nontrivial induced pairing in the lowest band, we follow Ref.\onlinecite{Alicea2010} to rewrite the Hamiltonian using the band basis, and the resulting effective $p$-wave pairing $\Delta^{(p)}$ is given by
\begin{eqnarray}
\label{pair}
\Delta^{(p)}&=& \Delta^{(p)}_{s, \alpha}+ \Delta^{(p)}_{s, \beta}+ \Delta^{(p)}_{t},\\\nonumber
\Delta^{(p)}_{s, \alpha} &\approx& \frac{\alpha}{2\sqrt{V^2+\alpha^2}}\Delta_0, \nonumber\\
\Delta^{(p)}_{s, \beta} &\approx& \frac{V\beta}{2\sqrt{V^2+\alpha^2+\beta^2}\sqrt{V^2+\alpha^2}}\Delta_0, \nonumber\\
\Delta^{(p)}_{t} & \approx &-\frac{V}{2\sqrt{V^2+\alpha^2}}d_z(k_x). \nonumber\\
\end{eqnarray}

Notably, there are three $p$-wave pairing terms in Eq.\ref{pair} arising from different physical origins. We denote the effective $p$-wave pairing due to singlet Cooper pairs as $\Delta_s^{(p)}$, and the pairing due to equal-spin triplet Cooper pairs as $\Delta_t^{(p)}$. 

\begin{figure}
 	\centering
 	\includegraphics[width=3.5in]{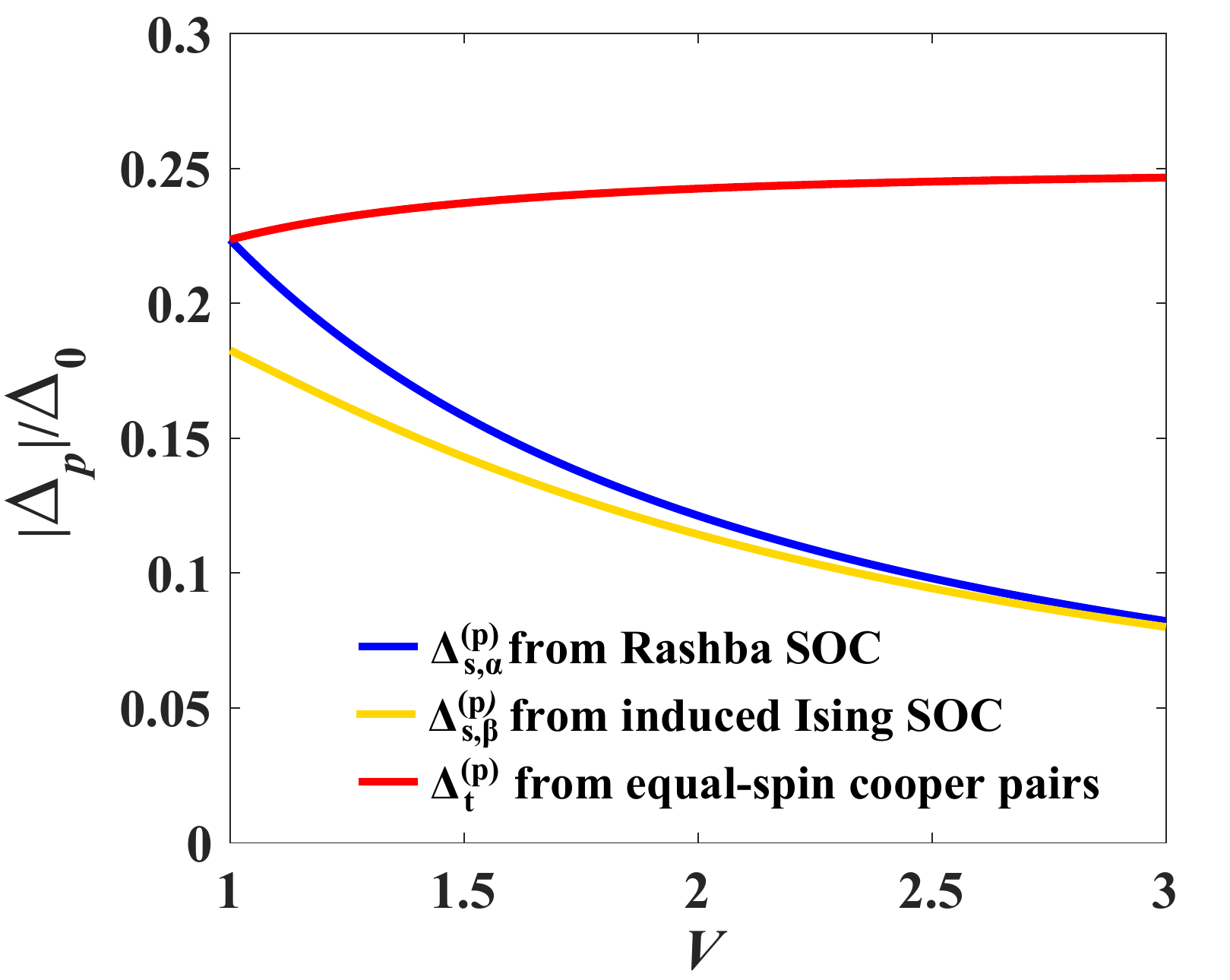}
 	\caption{Different $p$-wave pairing gaps as a function of Zeeman energy $V$. The blue/yellow curve represents $p$-wave pairing gap $\Delta_s^{(p)}$ induced by spin-singlet Cooper pairs and Rashba/induced Ising SOCs. Clearly, both of the induced gaps $\Delta_s^{(p)}$ are suppressed upon increasing $V$.  In contrast, the $p$-wave pairing gap $\Delta_t^{(p)}$ induced by equal-spin triplet Cooper pairs from Ising superconductors is insensitive to $V$. Here, we set the parameters to be $\alpha=0.5,\beta=0.5, d_z(k_F)/\Delta_0=0.5$.}
 	\label{fig:gapVsV}
 \end{figure}

The first $\Delta^{(p)}_{s, \alpha}$-term is the well studied effective $p$-wave pairing induced by Rashba SOC, Zeeman effect and proximity-induced $s$-wave pairing potential (blue curve in Fig.\ref{fig:gapVsV}), which is responsible for creating topological superconductivity in conventional InSb nanowires with Rashba SOCs. Here, $\alpha=\tilde{\mathcal{Z}}_0^{-1}\alpha_Rk_x/\Delta_0$ characterizes the strength of Rashba SOC strength in the nanowire, $V=\tilde{\mathcal{Z}}_0^{-1}V_x/\Delta_0$ is the renormalized Zeeman energy,  and $\Delta_0$ is the induced on-site $s$-wave component in $\psi(k_x)$. 

The second $\Delta^{(p)}_{s, \beta}$-term also results from proximity-induced singlet pairing(yellow curve in Fig.\ref{fig:gapVsV}), where $\beta=\beta_I(k_x)/\Delta_0$ is the proximity-induced Ising SOC in the nanowire. It arises from a similar physical mechanism as the $\Delta^{(p)}_{s, \alpha}$-term, except that the Ising SOC is induced in the nanowires from the background Ising superconductor. Notably, Eq.\ref{pair} indicates that both $\Delta^{(p)}_{s, \alpha}$ and $\Delta^{(p)}_{s, \beta}$ get strongly suppressed as the Zeeman effect dominates: $V \gg \alpha, \beta$.

In contrast, the last $\Delta_t^{(p)}$-term in Eq.\ref{pair} describes a triplet pairing which originates from the intrinsic equal-spin Cooper pairs in the parent Ising superconductor, with $d_z(k_x)\propto k_x$ (Appendix \ref{ApB}). It has a weak dependence on Zeeman energy $V$ and in the limit $V \gg \alpha,\beta$, we have $\Delta_t^{(p)} \approx -d_z(k_x)/2$. Thus,  $\Delta_t^{(p)}$ stays robust against Zeeman fields. 

To make an explicit comparison among all $p$-wave pairing terms above, we calculate the amplitude of each term as a function of Zeeman energy $V$(Fig.\ref{fig:gapVsV}). Evidently, both $\Delta^{(p)}_{s, \alpha}$ and $\Delta^{(p)}_{s, \beta}$ from intrinsic Rashba SOC/induced Ising SOC gradually decreases upon increasing $V_x$(blue/yellow curves in Fig.\ref{fig:gapVsV}), as the magnetic field tends to pin spins to the same directions and competes with the opposite spin-singlet pairing. In contrast, since $\Delta_t^{(p)}$ arises from equal-spin Cooper pairs with spins pointing to in-plane directions, it is compatible with in-plane fields and its amplitude remains almost unaffected by the in-plane field(red curve in Fig.\ref{fig:gapVsV}). 

%There are two more things about the results here that is necessary to be emphasised. First, we haven't considered the pairing gap of background superconductor changes with the in-plane magnetic field, which is true for Ising superconductor up to very high in-plane field. But the pairing gap of background will be quickly suppressed for the conventional SMN/SC setup because of the orbital effect and paramagnetic effect, for example,  NbTiN or epitaxial Al background either becomes soft gap or is totally suppressed at about only 2-3T\cite{Deng2016,Mourik2012}. So the decreasing of $\Delta_s^{(p)}$ will be much more fast comparing with Ising superconductor background. And in principle, there is a very robust $\Delta_t^{(p)}$ with the Ising superconducting background below the critical in-plane magnetic field which can be dozens of T. We will show this point in next part by performing a rather realistic numerical simulation. Second, it's very important to preserve a sizable $p$ wave gap to effectively realize the Kitaev chain so that the MFs are really well protected and robust against disorder. And preserving the topological superconducting gap up to stronger magnetic field also means a larger tunable topological region which lowers the difficulty of finning tunneling chemical potential and is very important for the scalable topological quantum computation. 

\subsection{Quasi-one-dimensional nanowires on superconducting atomically thin NbSe$_2$}\label{PartB}
In the previous subsection, we use a simplified strictly 1D model for InSb nanowires and a simple two-band model for Ising superconductors to illustrate the robust topological superconducting gap induced by equal-spin Cooper pairs. Here, we consider a realistic heterostructure formed by quasi-one dimensional nanowires and a specific Ising superconductor, monolayer 2H-NbSe$_2$\cite{Xiao2013}, to study the proximity-induced gap numerically. The set-up considered here is the same as in Fig.\ref{fig:nanotmd}.

\begin{figure}[ht]
	\centering
	\includegraphics[width=3 in]{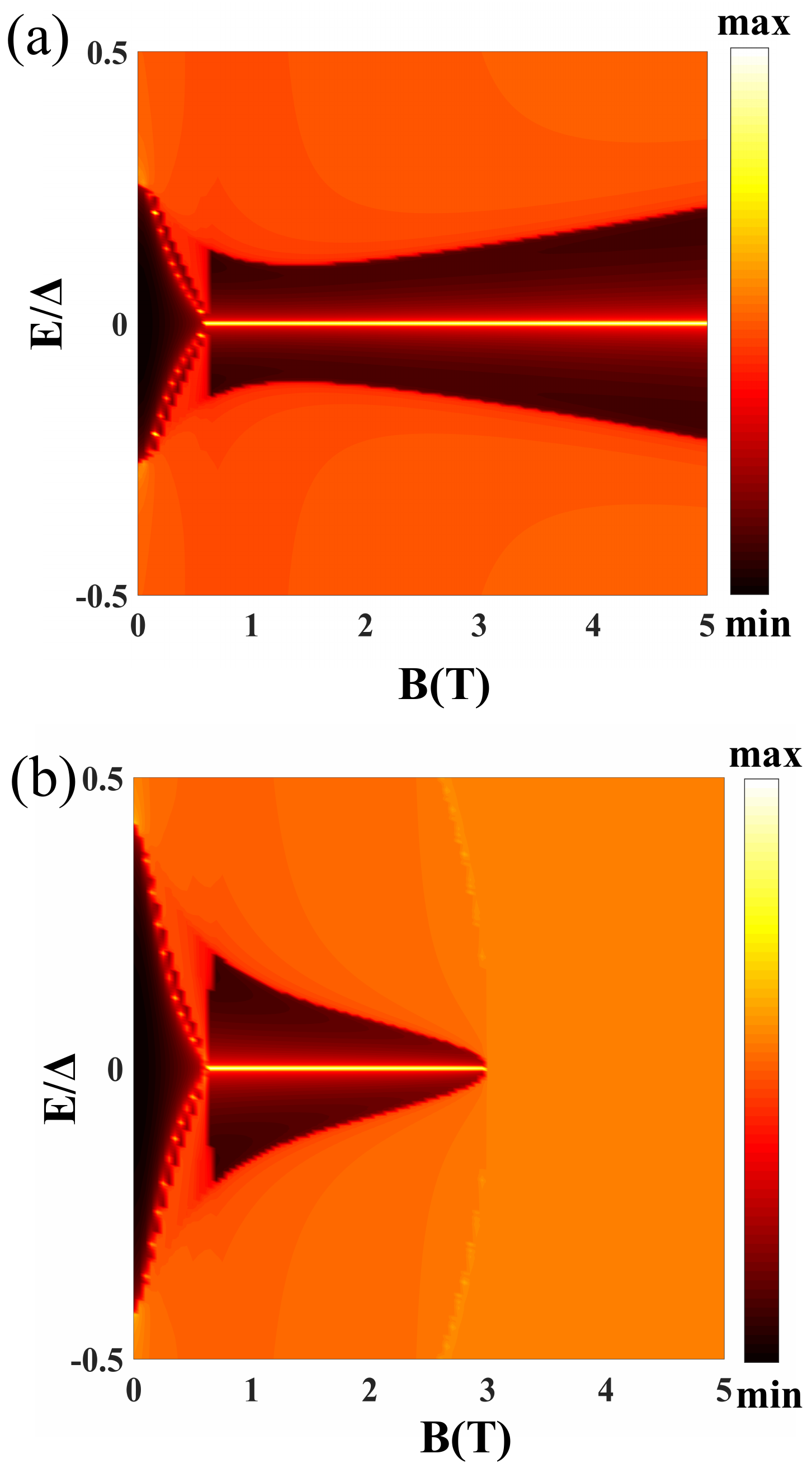}
%	\caption{A comparison between topological superconducting gaps induced by an Ising superconductor(a) and a conventional superconductor(b). (a) Local density of states at the end of InSb nanowires coupled to atomically thin superconducting NbSe$_2$. The topological gap shows a typical decaying behavior. We set the magnitude of superconducting gap as  $\Delta\sqrt{1-(B/B_c)^2}$ considering the  superconductivity will be gradually   suppressed in this case \cite{ChunxiaoLiu2017}. The critical field $B_c$ is set to be 3T. The chemical potential $\mu_w$ is chosen as 554.4 meV. (b) shows the local density states at the end of wire with Ising superconducting background, which is simulated with a superconducting monolayered NbSe$_2$ model $H_{TMD}$. The topological gap in this case is quite robust against the increasing of magnetic field.   The value of $\beta_{so}$ is taken form  the first principle calculation results in Ref.~(\onlinecite{He2018nodal}) as 78.4 meV.  And  $\mu_w=553.7$ meV.   }
	\caption{Comparison between topological superconducting gaps in (a)InSb nanowires/NbSe$_2$ heterostructure and (b)InSb nanowires/superconducting Al heterostructure. (a) Local density of states at the end of InSb nanowires coupled to atomically thin superconducting NbSe$_2$. The topological gap remains large at in-plane fields of a few Teslas, and persists up to conventional Pauli limiting fields $B \sim 10$T\cite{He2018}.  The value of $\beta_{so}$ is taken from  first-principle calculations in Ref.\onlinecite{He2018nodal}. (b) Local density states at the end of wire on top of a conventional superconductor without Ising SOC. The topological gap shows a typical decaying behavior. We set the magnitude of the parent superconducting gap as $\Delta\sqrt{1-(B/B_c)^2}$\cite{ChunxiaoLiu2017}. The critical field $B_c$ is set to be 3T.   }
	\label{fig:nbsenanowire}
\end{figure}

The realistic tight-binding model for monolayer 2H-NbSe$_2$\cite{Xiao2013} is
\begin{align}\label{TMD}
H_{TMD}&=\sum_{\bm{R},\alpha}(\epsilon_{\alpha}-\mu)\psi^{\dagger}_{\alpha,s}(\bm{R})\psi_{\alpha,s}(\bm{R})\nonumber\\
&+\sum_{\bm{R},\bm{d},s,\alpha,\beta}t_{\alpha\beta}(\bm{d})\psi^{\dagger}_{\alpha,s}(\bm{R})\psi_{\beta,s}(\bm{R}+\bm{d})+h.c.\nonumber\\
&+\sum_{\bm{R},s,s',\alpha,\beta}\beta_{so}\psi^{\dagger}_{\alpha ,s}(\bm{R})(\sigma_z)_{ss'}(L_z)_{\alpha\beta}\psi_{\beta,s'}(\bm{R})\nonumber\\
&+\sum_{\bm{R},s,s',\alpha}\Delta\psi^{\dagger}_{\alpha ,s}(\bm{R})(i\sigma_y)_{ss'}\psi^{\dagger}_{\alpha,s'}(\bm{R})+h.c..
\end{align}
The tight binding Hamiltonian for semiconductor nanowire is
\begin{align}\label{wire}
H_w=&\sum_{\bm{R},s,s'}c^{\dagger}_{s}(\bm{R})((4t_w-\mu_w)\delta_{ss'}+V_x(\sigma_x)_{ss'})c_{s'}(\bm{R})\nonumber\\
	&+\sum_{\bm{R},\bm{d},s}-t_wc^{\dagger}_{s}(\bm{R})c_{s}(\bm{R}+\bm{d})+h.c.+\sum_{\bm{R},\bm{d},s,s'}\frac{i}{2}\alpha_R\nonumber\\&c^{\dagger}_{s}(\bm{R})
	\hat{\bm{e}}_z\cdot(\bm{\sigma}_{ss'}\times\bm{d})c_{s'}(\bm{R}+\bm{d})+h.c..
\end{align}
The tunneling Hamiltonian is
\begin{equation}
H_c=\sum_{\bm{R}\in \text{wire},\alpha,s}\Gamma_c\psi^{\dagger}_{\alpha,s}(\bm{R})c_s(\bm{R})+h.c..
\end{equation}
The total Hamiltonian is
\begin{equation}
H_{t}=H_{TMD}+H_w+H_c,
\end{equation}
where $\alpha,\beta$ label different orbital, $s,s'$ label the spin, Pauli matrix $\sigma$  is defined in spin space, $L_z$ is the angular momentum operator defined in orbital space, $\epsilon_{\alpha}$ is the onsite energy for orbital $\alpha$, $\bm{d}$ is a lattice vector connecting the nearest and next nearest sites  of TMD or the nearest sites  of nanowire, $t_{\alpha\beta}(\bm{d}),t_w$ are the hopping strength, $\mu,\mu_w$ is the chemical potential of TMD and nanowire, $\beta_{so}$ is the strength of Ising SOC, $\alpha_R$ is the  Rashba SOC parameter, $\Gamma_c$ is the coupling strength between TMD and nanowire, $\psi_{\alpha,s}(\bm{R})$ and $c_{s}(\bm{R})$ are annihilation operators of the TMD and the nanowire.
\begin{table}
	\caption{\label{parameter} Parameters for tight-binding model estimated from Ref.\onlinecite{Mourik2012,Xi2015}. Parameters for 2H-NbSe$_2$ are adapted from \textit{ab initio}-based calculations in Ref.\onlinecite{He2018nodal}. }
	\begin{ruledtabular}
		\begin{tabular}{ccc}
			Parameter& Symbol & Value\\
			\hline
			The lattice constant of wire & a & 6 \AA{}\\
			Rashba parameter& $\alpha_R$&0.2 eV$\cdot$\AA{}\\
			Hopping strength of wire& $t_w$&250 eV$\cdot$\AA{}$^2$\\
			The magnetic energy&$V_x$&1.5 meV/T$\times B$\\
			Superconducting gap&$\Delta$&0.5 meV\\
			Coupling strength&$\Gamma_c$&120$\Delta$
		\end{tabular}
	\end{ruledtabular}
\end{table}  

In order to see how the pairing gap changes with magnetic field, we use realistic parameters of InSb nanowires\cite{Mourik2012,Lutchyn2018} and monalayer 2H-NbSe$_2$\cite{He2018nodal,Xiao2013} to calculate the local density of states(LDOS) at the end of the wire as a function of $V_x$(Fig.\ref{fig:nbsenanowire}). For a straightforward comparison, we model a usual $s$-wave superconductor by setting the Ising SOC to be zero in the NbSe$_2$. Details of these parameters are shown in Table \ref{parameter}. 

The LDOS at one end of InSb nanowires coupled to superconducting monolayer NbSe$_2$ is shown in Fig.\ref{fig:nbsenanowire}(a). Clearly, upon increasing magnetic field, the bulk excitation gap closes and reopens at $B \approx 0.5$T. The system enters the topologically nontrivial regime at the gap closing point, signified by the appearance of Majorana zero modes. Consistent with Fig.\ref{fig:gapVsV}, the topological superconducting gap of the InSb wire remains sizable for field strengths up to $B = 5$T according to our numerical calculations in Fig.\ref{fig:nbsenanowire}(a). Notably, for magnetic fields larger than $5$T, the wire still remains a topological superconductor as long as the proximity gap is finite. The proximity gap can eventually be destroyed if the parent superconducting gap in NbSe$_2$ is closed by the applied magnetic field. This can indeed happen at a field strength $B \sim 10$T corresponding to the conventional Pauli limit, where the parent superconducting NbSe$_2$ becomes a nodal topological superconductor\cite{He2018}. Therefore, the topological regime of the InSb wire is significantly enlarged with $B$-field ranging from $0.5- 10$T.

On the contrary, as shown in Fig.\ref{fig:nbsenanowire}(b), for a conventional parent superconductor with spin-singlet Cooper pairs only, the topological superconducting gap is suppressed quickly. To illustrate the differences between the InSb nanowire/NbSe$_2$ heterostructure and the InSb nanowire/superconductor(Al) heterostructures under realistic experimental conditions, in Fig.\ref{fig:nbsenanowire}(b) we also considered the orbital pair breaking effects, which was found to cause a magnetic field dependence of parent gap in superconducting aluminum: $\Delta(B) = \Delta \sqrt{1-(B/B_c)^2}$ with $B_c \approx 3$ T\cite{Marcus2016,ChunxiaoLiu2017}. As we mentioned earlier, the orbital pair breaking effect is negligible in atomically thin NbSe$_2$ as its thickness is far smaller than the superconducting coherence length. Thus, we conclude that an Ising superconductor such as atomically thin NbSe$_2$ has an overall advantage in inducing robust topological gap in InSb nanowires and strongly enlarged topological regime under external magnetic fields.

\section{Conclusion}

In conclusion, we showed that Ising superconductors such as atomically thin superconducting NbSe$2$ can induce robust topological superconducting gap in InSb nanowires with strongly enlarged topological regime. We explained that the robust proximity gap originates from the special equal-spin Cooper pairs in Ising superconductors which are compatible with in-plane magnetic fields. This robust topological superconducting gap with wide topological regime in Majorana nanowires induced by Ising superconductors provides a promising platform for robust Majorana-based qubits.

\begin{acknowledgments}
The authors thank Noah F. Yuan for helpful discussions. KTL acknowledges the support of Croucher Foundation, Dr. Tai-chin Lo Foundation and HKRGC through C6026-16W,
16324216, 16307117 and 16309718. \\
\end{acknowledgments}

\begin{appendix}
	
\section{Paring correlation from the Gor'kov equation}\label{Ap_Anomalous Green}
By solving the Gor'kov equation, we obtain
\begin{align}
	%&G(\bm{k},\bm{V},i\omega_n)=((\xi(\bm{k})-i\omega)|\bm{g}(-\bm{k},\bm{V})|-(\xi(\bm{k})+i\omega)(\xi^2(\bm{k})\nonumber\\&+\omega^2+\Delta^2))/(\varphi_{+}\varphi_{-})+(\Delta^2\bm{g}(-\bm{k},\bm{V})\nonumber+(|\bm{g}(-\bm{k},\bm{V})|^2\\
	%&-(\xi(\bm{k})+i\omega_n)^2))\bm{g}(\bm{k},\bm{V}))\cdot\bm{\sigma}/(\varphi_{+}\varphi_{-})\\
	&F_s(\bm{k},\bm{V},i\omega_n)=\frac{1}{2}(\frac{1}{\varphi_{+}(\bm{k},\bm{V},\Delta,\omega_n)}+\frac{1}{\varphi_{-}(\bm{k},\bm{V},\Delta,\omega_n)})\label{singletpaircorrelation},\\
	&\bm{F}_{t}(\bm{k},\bm{V},i\omega_n)=(-(\xi(\bm{k})+i\omega_n)\bm{g}(\bm{k},\bm{V})+(\xi(\bm{k})-\nonumber\\
	&i\omega_n)\bm{g}(-\bm{k},\bm{V})-i\bm{g}(\bm{k},\bm{V})\times\bm{g}(-\bm{k},\bm{V}))/(\varphi_{+}\varphi_{-}),
\end{align}
where $\varphi_{+}\varphi_{-}=\varphi_{+}(\bm{k},\bm{V},\Delta,\omega_n)\varphi_{-}(\bm{k},\bm{V},\Delta,\omega_n) ,\\\varphi_{\pm}(\bm{k},\bm{V},\Delta,\omega_n)=\Delta^2+\omega_n^2-|\bm{V}|^2+|\bm{g}(\bm{k})|^2+\xi^2(\bm{k})\pm2\sqrt{|\bm{g}(\bm{k})|^2\xi^2(\bm{k})-|\bm{V}|^2\omega_n^2+2i\bm{g}(\bm{k})\cdot\bm{V}\xi(\bm{k})\omega_n-|\bm{g}(\bm{k})\times\bm{V}|^2}$  and  $F(\bm{k},\bm{V},i\omega_n)$ has been parametrized as spin singlet pairing correlations $F_s(\bm{k},\bm{V},i\omega_n)$ and spin triplet pairing correlations $\bm{F}_t(\bm{k},\bm{V},i\omega_n)$, where $F(\bm{k},\bm{V},i\omega_n) =(F_s(\bm{k},\bm{V},i\omega_n)+\bm{F}_t(\bm{k},\bm{V},i\omega_n)\cdot\bm{\sigma})\Delta i\sigma_y$.         
	
%	\newpage
\section{Self-consistent gap equation for Ising superconductors}\label{ApA}
The self-consistent gap equation for Ising superconductor under an in-plane magnetic field is given by
\begin{equation}\label{self}
\Delta i\sigma_y=T\sum_{\bm{k},n}V_0F_s(\bm{k},V_x,i\omega_n)	\Delta i\sigma_y.
\end{equation}
The nontrivial solution $\Delta$ is obtained by solving Eq.\ref{self} self-consistently. Substitute Eq.\ref{singletpaircorrelation}, we obtain
\begin{align}
	1&=\frac{TV_0}{2}\sum_{\bm{k},n}(1-\frac{V_x^2}{\lambda(\bm{k})})\frac{1}{\omega_n^2+\chi^2_{-}(\bm{k})}\nonumber\\
	&+(1+\frac{V_x^2}{\lambda(\bm{k})})\frac{1}{\omega_n^2+\chi_{+}(\bm{k})^2},
\end{align}
where $\chi_{\pm}(\bm{k})=\sqrt{V_x^2+\beta^2(\bm{k})+\Delta^2+\xi^2(\bm{k})\pm2\lambda(\bm{k})}$, $\lambda(\bm{k})=\sqrt{V_x^2(\Delta^2+\xi^2(\bm{k}))+\beta^2(\bm{k})\xi^2(\bm{k})}$. Then we sum over the Matsubara frequencies
\begin{align}
		1&=\frac{V_0}{2}\sum_{\bm{k}}(1-\frac{V_x^2}{\lambda(\bm{k})})\frac{1}{2\chi_-(\bm{k})}\tanh(\frac{\chi_-(\bm{k})}{2T})\nonumber\\
	&+(1+\frac{V_x^2}{\lambda(\bm{k})})\frac{1}{2\chi_+(\bm{k})}\tanh(\frac{\chi_+(\bm{k})}{2T}).
\end{align}

By replacing $\sum_{\bm{k}}$ with $\int_{-\Lambda}^{\Lambda}d\xi \mathcal{N}(\mu_F)$($\Lambda$: cut-off energy), the self-consistent equation becomes
\begin{widetext}
\begin{align}\label{selfconsistent}
		\eta=&\int^{\sinh(\eta)/\delta}_{0}dx(1-\frac{\nu^2}{\sqrt{\nu^2\delta^2(1+x^2)+b^2\delta^2x^2}})\frac{\delta}{2\sqrt{\nu^2+b^2+\delta^2(1+x^2)-2\sqrt{\nu^2\delta^2(1+x^2)+b^2\delta^2x^2}}}\nonumber\\
		&\tanh(\frac{1.764\sqrt{\nu^2+b^2+\delta^2(1+x^2)-2\sqrt{\nu^2\delta^2(1+x^2)+b^2\delta^2x^2}}}{2\tau})+(1+\frac{\nu^2}{\sqrt{\nu^2\delta^2(1+x^2)+b^2\delta^2x^2}})\nonumber\\&\frac{\delta}{2\sqrt{\nu^2+b^2+\delta^2(1+x^2)+2\sqrt{\nu^2\delta^2(1+x^2)+b^2\delta^2x^2}}}\tanh(\frac{1.764\sqrt{\nu^2+b^2+\delta^2(1+x^2)+2\sqrt{\nu^2\delta^2(1+x^2)+b^2\delta^2x^2}}}{2\tau}),
\end{align}
\end{widetext}
where $\eta=\frac{1}{V_0\mathcal{N}(\mu_F)}, b=\beta_F/\Delta_0, \delta=\Delta(V_x)/\Delta_0, \nu=V_x/\Delta_0, \tau=T/T_c,x=\xi/\Delta(V_x)$.
$\mathcal{N}(\mu_F),\beta_F$ is density of states and the spin-orbit coupling strength near Fermi energy, $\Delta_0=1.764 T_c$ is the pairing gap at zero temperature without magnetic field, with $T_c$ being the critical temperature. Note that the effect of SOCs on $T_c$ is negligible\cite{Sigrist2004}.
 
Note that the gap equation only gives the saddle point of the free energy. To determine the pairing gap and the phase transition point, one needs to further compare the superconducting free energy $\mathcal{F}_s$ and the normal-state free energy $\mathcal{F}_n$. One way is to work out $\mathcal{F}_s-\mathcal{F}_n$ and evaluate $\int_0^{\Delta} d\Delta \frac{d(1/V_0)}{d\Delta} \Delta^2$ as done in Ref.\onlinecite{maki1964pauli}, where the interacting strength $V_0$ and pairing gap $\Delta$ are given by the self-consistent gap equation. 

However, due to the presence of both SOCs and magnetic field, the gap equation gets too involved to be solved exactly. Instead, we make an estimation by including the leading-order magnetization energy and condensation energy in $\mathcal{F}_s$ and $\mathcal{F}_n$, which leads us to the following form: $F_s-F_n\approx-\frac{1}{2}\chi_sB^2-\frac{1}{2}N(\mu_F)\Delta^2+\frac{1}{2}\chi_nB^2$, where $\chi_s,\chi_n$ is superconducting and normal spin susceptibility, $B$ is the in-plane magnetic field. In this way, the upper critical field $B_{c2}$ can be estimated as $u_BB_{c2}\approx\frac{\Delta}{\sqrt{2}\sqrt{1-\chi_s/\chi_n}}$\cite{PAFrigeri}. For the Ising superconductor, the superconducting spin susceptibility is $\chi_s=\chi_n(1-\pi k_BT\sum_{n}\frac{1}{\omega_n^2+\Delta^2+\beta^2(\mu_F)}\frac{\Delta^2}{\sqrt{\omega_n^2+\Delta^2}})$ according to Ref.\onlinecite{PAFrigeri}. Combining with self-consistent gap equation, we can obtain $\chi_s/\chi_n$ at zero temperature as shown in  Fig.\ref{fig:gapv}(a). We see that $\chi_s/\chi_n$ increases with the Ising SOC strength and approaches to 1. This explains the enhancement of in-plane $B_{c2}$ shown in Fig.\ref{fig:gapv}(b).
 
\section{The low-energy effective Hamiltonian of nanowires on Ising superconductors}\label{ApB}
The Green's function of a bulk Ising superconductor is
\begin{equation}
\mathcal{G}(\bm{k},i\omega)=
\begin{pmatrix}
G(\bm{k},\bm{0},i\omega)&-F(\bm{k},\bm{0},i\omega)\\
-F^{\dagger}(\bm{k},\bm{0},i\omega)&-G^{T}(-\bm{k},\bm{0},-i\omega)
\end{pmatrix}.
\end{equation}
Here we parameterize $G(\bm{k},i\omega)$ which can be obtained from the Gor'kov equations as
	 $G_{+}(\bm{k},i\omega)+G_{-}(\bm{k},i\omega)\hat{\bm{g}}(\bm{k})\cdot\bm{\sigma}$. And 
	 \begin{eqnarray}
	 G_{+}(\xi(\bm{k}),i\omega)&=-\frac{1}{2}(\frac{i\omega+\xi_{+}(\bm{k})}{\varphi_{+}(\bm{k},\bm{0},\Delta,\omega)}+\frac{i\omega+\xi_{-}(\bm{k})}{\varphi_{-}(\bm{k},\bm{0},\Delta,\omega)}),\\
	  G_{-}(\xi(\bm{k}),i\omega)&=-\frac{1}{2}(\frac{i\omega+\xi_{+}(\bm{k})}{\varphi_{+}(\bm{k},\bm{0},\Delta,\omega)}-\frac{i\omega+\xi_{-}(\bm{k})}{\varphi_{-}(\bm{k},\bm{0},\Delta,\omega)}).
	 \end{eqnarray}
By integrating out the superconducting background, the Green's function of nanaowire is
\begin{eqnarray}
G^{-1}_{w}(k_x,i\omega)=i\omega-H_{w}(k_x)-\Sigma(k_x;i\omega).
\end{eqnarray}
Here, $\Sigma(k_x;i\omega) = \int \frac{d\bm{k}_{\perp}}{(2\pi)^{d-1}}\mathcal{T}^{\dagger}\mathcal{G}(\bm{k},i\omega)\mathcal{T}$ is the self-energy of the Ising superconductor. Then we expand $G^{-1}_{w}(k_x,i\omega)$ in the linear order of $\omega$ such that $G^{-1}_{w}(k_x,\omega)=\tilde{\mathcal{Z}}_0(i\omega-\hat{H}_{eff}(k_x))$. The low-energy effective Hamiltonian is thus given by
	\begin{eqnarray}
	\hat{H}_{eff}(k_x)=\begin{pmatrix}
	\tilde{h}(k_{x})&\tilde{\Delta}(k_x)\\
\tilde{\Delta}(k_x)^{\dagger}&-\tilde{h}^{T}(-k_{x})
	\end{pmatrix}.
	\end{eqnarray}
Here the normal part, with the wire Hamiltonian $h_{w}(k_{x})=\frac{\hbar^2k_{x}^2}{2m}-\mu_w+\alpha_Rk_x\sigma_y+V_x\sigma_x$, is given by: 
\begin{eqnarray}
\tilde{h}(k_{x})&=&\frac{h_{w}(k_{x})}{\tilde{\mathcal{Z}_0}}-\frac{\Gamma_c^2}{2\tilde{\mathcal{Z}_0}}\int \frac{d\bm{k}_{\perp}}{2\pi}(\frac{\xi_{+}(\bm{k})}{\xi^2_{+}(\bm{k})+\Delta^2}+\frac{\xi_{-}(\bm{k})}{\xi^2_{-}(\bm{k})+\Delta^2})\nonumber\label{dvector}\\
&-&\frac{\Gamma_c^2}{2\tilde{\mathcal{Z}_0}}\int \frac{d\bm{k}_{\perp}}{2\pi}(\frac{\xi_{+}(\bm{k})}{\xi^2_{+}(\bm{k})+\Delta^2}-\frac{\xi_{-}(\bm{k})}{\xi^2_{-}(\bm{k})+\Delta^2})\hat{\bm{g}}(\bm{k})\cdot\bm{\sigma}.\nonumber\\
\end{eqnarray}
The induced pairing gap function is: $\tilde{\Delta}(k_x)=(\psi(k_{x})+\bm{d}(k_{x})\cdot\bm{\sigma})i\sigma_y$, where
\begin{eqnarray}
\psi(k_{x})&=&-\frac{\Delta\Gamma_c^2}{\tilde{\mathcal{Z}_0}}\int \frac{d\bm{k}_{\perp}}{2\pi}F_{s}(\bm{k},i\omega)\nonumber\\
&=&-\frac{\Delta\Gamma_c^2}{2\tilde{\mathcal{Z}_0}}\int \frac{d\bm{k}_{\perp}}{2\pi}(\frac{1}{\xi^2_{+}(\bm{k})+\Delta^2}+\frac{1}{\xi^2_{-}(\bm{k})+\Delta^2}),\nonumber\\
\bm{d}(k_{x})&=&-\frac{\Delta\Gamma_c^2}{\tilde{\mathcal{Z}_0}}\int \frac{d\bm{k}_{\perp}}{2\pi}\bm{F_{t}}(\bm{k},i\omega)\nonumber\\
&=&-\frac{\Delta\Gamma_c^2}{2\tilde{\mathcal{Z}_0}}\int \frac{d\bm{k}_{\perp}}{2\pi}(\frac{1}{\xi^2_{+}(\bm{k})+\Delta^2}-\frac{1}{\xi^2_{-}(\bm{k})+\Delta^2})\hat{\bm{g}}(\bm{k})\label{spin triplet},\nonumber\\
\tilde{\mathcal{Z}_0}&=&1+\frac{\Gamma_c^2}{2}\int \frac{d\bm{k}_{\perp}}{2\pi}(\frac{1}{\xi^2_{+}(\bm{k})+\Delta^2}+\frac{1}{\xi^2_{-}(\bm{k})+\Delta^2})\label{normal},
\end{eqnarray}
For an Ising superconductor, we have $\bm{g}(\bm{k})=\beta(\bm{k}) \hat{\bm{e}}_z$. The induced SOC and $\bm{d}$-vector are
\begin{eqnarray}
\beta_I(k_x)&=&\frac{\Gamma_c^2}{\tilde{2\mathcal{Z}_0}}\int \frac{d\bm{k}_{\perp}}{2\pi}[\frac{-\xi_{+}(\bm{k})}{\xi^2_{+}(\bm{k})+\Delta^2}+\frac{\xi_{-}(\bm{k})}{\xi^2_{-}(\bm{k})+\Delta^2}], \\\nonumber
\bm{d}(k_x)&=&\frac{\Delta\Gamma_c^2}{2\tilde{\mathcal{Z}_0}}\int \frac{d\bm{k}_{\perp}}{2\pi}[\frac{1}{\xi^2_{+}(\bm{k})+\Delta^2}-\frac{1}{\xi^2_{-}(\bm{k})+\Delta^2}]\hat{\bm{e}}_z\label{dz}.
\end{eqnarray}	
The above results can be further simplified by rewriting
\begin{eqnarray}
F(k_x)&=&\int\frac{d\bm{k_\perp}}{2\pi}\frac{f(\bm{k}_{\perp},k_x)}{\xi_{\pm}^2(\bm{k}_{\perp},k_x)+\Delta^2}\nonumber\\
%&=&\int \frac{d\bm{k_\perp}}{(2\pi)^{d-1}}\int_{-\Lambda}^{\Lambda}\frac{d\epsilon}{\epsilon^2+\Delta^2}\delta(\epsilon-\xi_{\pm}(\bm{k}))f(\bm{k}_{\perp},k_x)\nonumber\\
%&=&\int \frac{d\bm{k_\perp}}{(2\pi)^{d-1}}\frac{2\pi i}{2\Delta i}\delta(\Delta i-\xi_{\pm}(\bm{k}))f(\bm{k}_{\perp},k_x)\nonumber\\
&\approx&\frac{\pi}{\Delta}\left\langle\rho_{\pm}(\bm{k}_\perp) f(\bm{k_\perp},k_x)\right\rangle_{|\bm{k}_\perp|^2=k_F^2-k_x^2},
\end{eqnarray} 

where the density of states in a unit length of $\bm{k}_\perp$  is $ \rho_{\pm}(\bm{k}_\perp)= |\bm{\nabla_{k_\perp}}\xi_{\pm}(\bm{k})|^{-1}$, the cut off energy $\Lambda\gg \Delta$, and $\left\langle \cdots\right\rangle_{|\bm{k}_\perp|^2=k_F^2-k_x^2}$ denotes average over the Fermi circles. The final results are
\begin{eqnarray}
  \psi(k_x)&=&\frac{-\pi\Delta\Gamma_c^2\sum_{\lambda=\pm}\left\langle \rho_{\lambda}(\bm{k}) \right\rangle_{|\bm{k}_\perp|^2=k_F^2-k_x^2}}{2\Delta+\pi\Gamma_c^2\sum_{\lambda=\pm}\left\langle \rho_{\lambda}(\bm{k}) \right\rangle_{|\bm{k}_\perp|^2=k_F^2-k_x^2}}\label{sinduce},\\\nonumber
  \bm{d}(k_x)&=&\frac{-\pi\Delta\Gamma_c^2\left\langle \rho_{+}(\bm{k})-\rho_{-}(\bm{k}) \right\rangle_{|\bm{k}_\perp|^2=k_F^2-k_x^2}}{2\Delta+\pi\Gamma_c^2\sum_{\lambda=\pm}\left\langle \rho_{\lambda}(\bm{k}) \right\rangle_{|\bm{k}_\perp|^2=k_F^2-k_x^2}}\bm{\hat{e}}_z.
\end{eqnarray}

The $D_{3h}$ point group of a monolayer 2H-NbSe$_2$ dictates the SOC of the form $\beta(\bm{k}) = \beta_{so}(\sin(k_x)-2\sin(k_x/2)\cos(\sqrt{3}/2k_y))$\cite{bauer2012non}. Therefore, near the $\Gamma$-point where the topological phase transition happens, $\psi(k_x)\approx \Delta_0,d_z(k_x)\propto k_x$, namely, $\psi(k_x)$ reduces to an $s$-wave-like gap, but $d_z(k_x)$ becomes a $p$-wave-like gap. 
\end{appendix}
\bibliographystyle{apsrev4-1} 
\bibliography{Ref}
%\bibitem [{\citenamefont {Ivanov}(2001)}]{PhysRevLett.86.268}

\end{document}